\documentstyle[12pt]{article}
\begin{document}

\begin{titlepage}

\begin{flushright}
CIEA-GR-9402
\end{flushright}
\begin{center}
\vskip 3em
{\Large No Quantum Super-minisuperspace with $\Lambda \ne 0$}
\vskip 2em
{\large Riccardo Capovilla${}^{(a)}$ and
 Octavio Obreg\'on${}^{(b)}$} \\[1em]
\em{${}^{(a)}$ Departamento de F\'{\i}sica \\
Centro de Investigaci\'on y de Estudios Avanzados, I.P.N.\\
Apdo. Postal 14-740, M\'exico 14, D.F., M\'exico.\\[.7em]
${}^{(b)}$ Instituto de F\'{\i}sica\\
Universidad de Guanajuato \\
Apdo. Postal E-143, 37150 Le\'on, Gto., M\'exico. \\
and \\
Departamento de F\'{\i}sica\\
Universidad Aut\'onoma Metropolitana,
Itzapalapa\\
Apdo.Postal 55-534, 09340 D.F., M\'exico}
\end{center}
\vskip 1em
\begin{abstract}
We show that the quantum super-minisuperspace
of N=1 supergravity with $\Lambda \ne 0 $ has no non-trivial
physical states for class A Bianchi models.
Hence, in super quantum cosmology, the vanishing of
$\Lambda$ is a condition for the existence of the
universe.
We argue that this result implies that in full supergravity
with $\Lambda$
there are no non-trivial physical states
with a finite number of fermionic fields.
We use the Jacobson canonical formulation.
\end{abstract}
\end{titlepage}
\newpage


\def\eg{{\it e.g.${~}$}}
\def\ie{{\it i.e.${~}$}}
\def\frac#1#2{{\textstyle{#1\over\vphantom2\smash{\raise.20ex
        \hbox{$\scriptstyle{#2}$}}}}}
\def\st{\widetilde{\sigma}}
\def\pt{\widetilde{\pi}}
\def\e{\varepsilon}
\def\D{{\cal{D}}}
\def\p{\psi}
\def\s{\sigma}
\def\H{{\cal{H}}}
\def\J{{\cal{J}}}
\def\Sc{{\cal{S}}}
\def\Sh{{\cal{S}}^{\dagger A}}
\def\ph{\psi^{\dagger}}
\def\Rc{{\cal{R}}}
\def\Rh{{\cal{R}}^{\dagger A}}


\def\ga{\mbox{\bf {\scriptsize{a}}}}
\def\gb{\mbox{\bf {\scriptsize{b}}}}
\def\gc{\mbox{\bf {\scriptsize{c}}}}
\def\gd{\mbox{\bf {\scriptsize{d}}}}
\def\ge{\mbox{\bf {\scriptsize{e}}}}
\def\gf{\mbox{\bf {\scriptsize{f}}}}
\def\gg{\mbox{\bf {\scriptsize{g}}}}
\def\gh{\mbox{\bf {\scriptsize{h}}}}
\def\gi{\mbox{\bf {\scriptsize{i}}}}
\def\gj{\mbox{\bf {\scriptsize{j}}}}


\noindent{\bf I. Introduction}

\vskip .5em

In a recent paper, the super-minisuperspace of
canonical N=1 supergravity in the formulation
given by Jacobson \cite{Jac} was introduced \cite{CG}.
The quantization of all Bianchi type A models was carried out,
in the triad representation.
Exploiting the fact that the system has a finite number of
degrees of freedom, the wavefunction can be expanded in (even) powers
of the gravitino field, up to sixth order.
It was found that, in general,
the physical states of the quantum theory have a very restricted
form. Only the parts of the wavefunction
 of zero and sixth order in the gravitino field are non-vanishing.

In this paper, we point out that if one includes
a cosmological constant $\Lambda$ in super-minisuperspace
(first considered
 in this context in \cite{Gra,MOS}),
 the only physical state allowed is trivial,
\ie $\Psi = 0$, for any factor ordering. Since we find
$\Psi = 0$, this will also be true in any representation.
The technical reason for this is that,
in the presence of $\Lambda$, the quantum supersymmetry constraints
mix parts  of the wavefunction of different order in the
 gravitino field. This mixing does not allow to fullfill
 the coupled system of quantum constraint equations, except with
the trivial solution.

{}From this result one arrives to the conclusion that,
if one is confident in the super-minisuperspace
approximation,
the price of introducing a cosmological constant in
super-minisuperspace is the universe itself!
To recall Einstein's words, to introduce a cosmological
constant would be a big blunder, indeed. Moreover,
we expect the same conclusion to hold if one adds
(supersymmetric) matter to this models, because the same mixing
of parts of the wavefunction of different order
in the gravitino fields will continue to take place.

The implications for the full theory are that in the
presence of a non-vanishing cosmological constant,
the only state with a {\it finite} number of fermionic
fields is trivial. We can offer two arguments for this.
The first is that the above mentioned mixing will still
be present. The
second argument is that if non-trivial states with a finite
number of fermionic fields were allowed in the full theory,
they would certainly show up in the mini-superspace models.
{}From these arguments, it follows that
physical states for supergravity with $\Lambda$ contain
an infinite number of fermionic fields.
(Carroll et al. reach the same conclusion for the
case $\Lambda = 0$ \cite{CFOP}. See, however, Ref. \cite{DEF} for
a different point of view.)

It is interesting to note that these quantum states are
not accessible in the super-minisuperspace approximation.
Therefore, our result can be considered as an explicit example
of the intrinsic limitations of the method of quantizing
the homogenous sector of the classical theory.

The paper is organized as follows. In Sect. II, we recall
briefly the necessary elements of canonical supergravity with
$\Lambda$, and we specialize to homogenous solutions.
In Sect. III, we quantize class A Bianchi models in the
triad representation. In Sect. IV, we give the explicit proof
that only $\Psi = 0$ solves the quantum supersymmetry constraints.
We conclude in Sect. V with some final remarks about previous work
in this topic.

\vskip 1em

\noindent {\bf II.  Super-minisuperspace with $\Lambda \ne 0$}

\vskip .5em

As canonical coordinates for N=1 supergravity we take a complex
traceless
$SL(2,C)$ spatial connection,
$A_{iA}{}^B$, and a traceless vector density of weight 1,
$\st^{iAB}$, together with the spatial
anti-commuting gravitino field $\p_i^A$, and its conjugate momentum
the anti-commuting
$\pt^{iA}$ [1,2]. (Small latin letters from the middle of the alphabet
denote spatial indices, $i,j,\dots = 1,2,3$. Capital latin letters
denote $SL(2,C)$ indices $A,B,\dots = 0,1$. These indices are raised
and lowered with the anti-symmetric symbol $\epsilon^{AB}$, and its
inverse $\epsilon_{AB}$, according to the rules $\lambda^A =
\epsilon^{AB} \lambda_B$, $\lambda_A = \lambda^B \epsilon_{BA}$.)

 The
connection $A_{iA}{}^B$ is the spatial pull-back of the
left-handed spin connection. The vector density
$\st^{iAB}$ may be interpreted as the (densitized) spatial
triad, in the sense that the covariant
(doubly densitized) spatial metric is given by
$ (det q) q^{ij} = \st^{i AB} \st^j{}_{AB} $.
The momentum $\pt^i_A $ is related to the complex conjugate of the
spatial gravitino field, $\psi_i^A$.

In the presence of a cosmological constant,
$\Lambda = -  4 m^2$, the supersymmetry constraints are given by
\cite{Jac,PKT}
\begin{eqnarray}
\Rc^A &:=& \Sc^A + \Sc_m^A = 0, \\
\Rh &:=& \Sh + \Sh_m = 0.
\end{eqnarray}
We write separately the parts that correspond to a vanishing
cosmological constant, given by,
\begin{eqnarray}
\Sc^A &:=& \D_k \pt^{kA},
\label{eq:susy1} \\
\Sh &:=& (\st^j \st^k \D_{[j} \p_{k]} )^A .
\label{eq:susy2}
\end{eqnarray}
Here $\D_i$ is the covariant derivative of $A_{iA}{}^B$.
We are
using the convention that
suppressed spinor indices are contracted from
upper left to lower right, \eg
$ ( \st^i \st^j )^{AB} = \st^{iAC} \st^j{}_C{}^B $.
Hermitian conjugation is defined with respect to some
Hermitian metric $n^{AA'}$.

The parts proportional to the cosmological constant are
\begin{eqnarray}
\Sc_m^A &:=& i 2 \sqrt{2} m (\st^i \psi_i )^A ,\\
\Sh_m &:=&   - i 2 \sqrt{2} m (\st_i \pt^i )^A .
\end{eqnarray}

There are also constraints corresponding to the additional
invariance of the theory, \ie $SL(2,C)$ and diffeomorphism
invariance, but we will not need their explicit form here
(see [1]).

In the case of spatial homogeneity, we consider a
kinematical triad of vectors, $X^i_{\ga}$, which commute
with the three Killing vectors on the spatial hypersurface
\cite{AP}.  (Latin letters from the
beginning of the alphabet, $a,b,c,...$
label the triad vectors.)
The triad satisfies
$
[ X_{\ga} , X_{\gb} ]^i  = C_{\ga \gb}{}^{\gc}  X^i_{\gc} \; ,
$
where $C_{\ga \gb}{}^{\gc} $ denote the structure constants
of the Bianchi type under consideration. The basis dual
to $X^i_{\ga}$ is defined with
$ X^i_{\ga} \; \; \chi_i^{\gb} = \delta_{\ga}^{\gb}$.
We restrict to type A Bianchi models by assuming that
$C_{\ga \gb}{}^{\gc} = \epsilon_{\ga \gb \gd} M^{\gd \gc}$,
with $M^{\ga \gb}$ symmetric. (For our reasons to
exclude class B Bianchi models, see \cite{CG}.)

 The Jacobson phase space variables
may be expanded
with respect to the kinematical triad $X^i_{\ga}$ (or $\chi_i^{\gb}$), as,
\begin{eqnarray}
A_{i A}{}^B &=& A_{\ga A}{}^B \chi_i^{\ga}, \nonumber\\
\st^{iAB} &=& (det\chi )\; \s^{\ga AB} X^i_{\ga}, \nonumber \\
\p_i^A  &=& \p_{\ga}^A \chi_i^{\ga},  \nonumber \\
\pt^{iA} &=& (det\chi ) \pi^{\ga A} X^i_{\ga}, \nonumber
\end{eqnarray}
where we introduce $det \chi$ to de-densitize the momentum variables.

Inserting the expansion in the supersymmetry constraints (1) and (2),
for Bianchi class A models they reduce to
\begin{eqnarray}
R^{\dagger A} &=& S^{\dagger A} + S_m^{\dagger A}, \\
R^A &=& S^A + S_m^A ,
\end {eqnarray}
where
\begin{eqnarray}
S^{\dagger A} &=&  - {1 \over 2} \e_{\ga \gb \gd} M^{\gd \gc} (\s^{\ga}
\s^{\gb} \psi_{\gc } )^A
+ ( \s^{[\ga } \s^{\gb]} A_{\ga} \psi_{\gb} )^A, \\
S^A &=& A_{\ga}{}^{AB} \pi^{\ga}_B ,\\
S_m^{\dagger A} &=& - i 2 \sqrt{2} m h^{-1/2} ( \s_{\ga} \pi^{\ga})^A ,\\
S_m^A &=& i 2 \sqrt{2} m (\s^{\ga} \psi_{\ga} )^A .
\end{eqnarray}
We denote with $h$ the determinant of $h_{\ga \gb }
:= Tr (\s_{\ga} \s_{\gb} )$, and $\s_{\ga} $ is the inverse of
$\s^{\ga}$. Note that we have rescaled the constraints by
the appropriate factor of $det \chi$ to de-densitize them.

\vskip 1em

\noindent{\bf III. Quantization in the Triad Representation}

\vskip .5em

Now we turn to the quantization of class
A Bianchi models in the triad representation.
Quantum states may be
represented by wavefunctions that
depend on the triad, and on the gravitino field,
$\Psi = \Psi (\s, \psi)$.
The variables $\s^{\ga AB}$ and $\psi{\ga }^A$ turn into
 `position' operators. Their momenta
are represented  with
$
\hat{A}_{\ga}{}^{AB} \Psi =  (1/\sqrt{2})( \delta \Psi /
\delta \s^{\ga}{}_{AB} )
,
\hat{\pi}^{\ga}{}_A \Psi = (1/\sqrt{2}) (\delta \Psi /
\delta \psi_{\ga}{}^A )
$.

The translation of the supersymmetry constraints to their quantum version
gives
\begin{eqnarray}
\hat{R}^{\dagger A} \Psi &=& ( \hat{S}^{\dagger A} +
\hat{S}_m^{\dagger A} ) \Psi = 0 ,\\
\hat{R}^A \Psi &=& ( \hat{S}^A + \hat{S}_m^A )\Psi = 0,
\end {eqnarray}
where
\begin{eqnarray}
\hat{S}^{\dagger A} \Psi &=&
[ - {1 \over 2} \e_{\ga \gb \gd} M^{\gd \gc} (\s^{\ga}
\s^{\gb} \psi_{\gc })^A
+  {1  \over  \sqrt{2}} (\s^{[\ga} \s^{\gb]} {\delta \over \delta \s^{\ga} }
\psi_{\gb})^A
+ {1  \over  \sqrt{2}} \alpha \s^{\ga AC} \psi_{\ga C}] \Psi
\nonumber  \\
\hat{S}^A \Psi &=&  {1 \over 2} {\delta^2 \Psi
\over \delta \s^{\ga}{}_{AB} \delta \psi_{\ga}{}^B } \nonumber\\
\hat{S}_m^{\dagger A} \Psi &=&
- i 2  m h^{-1/2} ( \s_{\ga}{}
{\delta \Psi \over \delta \psi_{\ga}} )^A  \nonumber \\
 \hat{S}_m^A &=&   i 2 \sqrt{2} m (\s^{\ga} \psi_{\ga} )^A  \Psi.
\nonumber
\end{eqnarray}
The constant $\alpha $ parametrizes the factor ordering ambiguity in (9).

In addition,  a physical state must also satisfy
$\hat{J}^{AB} \Psi = 0 $, where $J^{AB} $ is the generator
of $SL(2,C)$ transformations. It follows
that the wavefunction must be an $SL(2,C)$ scalar.
Invariance under diffeomorphisms of a physical state
follows automatically from (13,14), because of the
well known fact that $\{ R^{\dagger A}, R^A \} \propto H^{AB} $,
where $H^{AB}$ are the diffeomorphism constraints.

Since the hamiltonian system is finite dimensional,
the wavefunction may be expanded in powers of the
anti-comuting gravitino fields.
The requirement of $SL(2,C)$ invariance
dictates that only terms of even power of the gravitino
field will appear. There are six components in $\psi_{\ga}{}^A$,
so the expansion stops at order six. Therefore we can write,
symbolically
\begin{equation}
\Psi (\s ,\psi ) =  \Psi_{(0)} (\s ) + \Psi_{(2)} (\s ,\psi ) +
\Psi_{(4)} (\s ,\psi ) + \Psi_{(6)} (\s ,\psi ),  \nonumber
\label{eq:wf}
\end{equation}
where the subscript indicates the number of gravitino fields.

When this decomposition is inserted in Eqs. (13, 14),
one finds the following
equations, that show explicitly the mixing,
\begin{eqnarray}
\hat{S}^{\dagger A} \Psi_{(n)} +
\hat{S}_m^{\dagger A}  \Psi_{(n+2)} &=& 0,
\label{eq:mx1}\\
\hat{S}^A \Psi_{(n+2)} + \hat{S}_m^A \Psi_{(n)} &=& 0,
\label{eq:mx2}
\end {eqnarray}
where $n = 0,2,4$.

At this point, it is convenient to
identify the irreducible
spin 1/2 and spin 3/2 parts of the gravitino field.
Let,
\begin{equation}
\psi^{ABC}  := \psi_{\ga}^A \s^{\ga BC} = \psi^{A(BC)}.
\end{equation}
Then,
\begin{equation}
\psi^{ABC} = \rho^{ABC} + \epsilon^{A(B} \beta^{C)}
\end{equation}
where $\rho^{ABC} = \rho^{(ABC)}$,
represents the spin 3/2 part, and
$\beta^A = (2/3) \psi_B{}^{AB}$, the spin 1/2 part.
It is important to note that they can be specified
independently.

The wavefunction can depend on the gravitino field
only in Lorentz invariant combinations. Thus,
the most general expression for the terms $\Psi_{(n)}$
in (\ref{eq:wf})
is
\begin{eqnarray}
\Psi_{(0)} &=& E(h), \nonumber \\
\Psi_{(2)} &=&  F_1 (h) [\rho^2 ] + F_2 (h) [\beta^2], \nonumber \\
\Psi_{(4)} &=& G_1 (h) [\rho^2]^2  + G_2 (h) [\rho^2] [\beta^2],
\nonumber\\
\Psi_{(6)} &=&  H (h) [\rho^2]^2 [\beta^2], \nonumber
\end{eqnarray}
where the functions $E,F,G,H$ depend on $\s^{\ga AB}$ only in the combination
$h^{\ga \gb } = Tr ( \s^{\ga} \s^{\gb}) $, and
$ [\rho^2 ] := \rho^{ABC} \rho_{ABC} ,
[\beta^2] := \beta^A \beta_A. $

\vskip 1em

\noindent{\bf IV. Physical States}

\vskip .5em

In this section we give the explicit proof
that all $\Psi_{(n)}$ terms
are zero, if they have to satisfy the quantum supersymmetric
constraints equations .

Our first step is to express the constraints
$\hat{S}^{\dagger A} $
 and $\hat {S}^{\dagger A}_{m} $ in terms of the independent quantities
 $\beta$ and $\rho$, acting on each term of the expansion
(\ref{eq:wf}). We find
\begin{eqnarray}
 \sqrt {2} \hat{S}^{\dagger A} \Psi_{(0)} &=&   \hat{T}^A (\rho) E
 + \beta^A [ h^{\ga \gb} {\delta E \over \delta h^{\ga \gb}}
 - {1 \over 2} h^{-1/2} M^{\ga \gb} h_{\ga \gb} E
+ {3 \over 2} \alpha E ],  \nonumber \\
 \sqrt {2} \hat{S}^{\dagger A} \Psi_{(2)} &=&  [\rho^2]  \hat{T}^A
 (\rho) F_1
 +  [\beta^2] \hat{T}^A (\rho) F_2
 + \beta^A [\rho^2 ] [ h^{\ga \gb}
{\delta F_1 \over \delta h^{\ga \gb}} \nonumber \\
 &+& {1 \over 6} F_2 + (  - {1 \over 2}
h^{-1/2} M^{\ga \gb} h_{\ga \gb}
 + {3 \over 2} \alpha
+ 2 )F_1 ], \nonumber \\
\sqrt {2} \hat{S}^{\dagger A} \Psi_{(4)} &=&  [\rho^2] [\beta^2]
 \hat{T}^A (\rho) G_2 +
 \beta^A [\rho^2 ]^2 [ h^{\ga \gb} {\delta G_1 \over \delta h^{\ga \gb}}
 + {1 \over 6} G_2 \nonumber \\
 & + &  (  - {1 \over 2} h^{-1/2} M^{\ga \gb} h_{\ga \gb}
+ {3 \over 2} \alpha + 2 ) G_1 ], \nonumber \\
 \hat{S}_m^{\dagger A}  \Psi_{(2)} &=&  4 i m h^{-1/2} \beta^A F_2 ,
\nonumber\\
\hat{S}_m^{\dagger A}  \Psi_{(4)} &=&  4 i m h^{-1/2} \beta^A [\rho^2]
 G_2 , \nonumber \\
\hat{S}_m^{\dagger A}  \Psi_{(6)} &=&
4 i m h^{-1/2} \beta^A [\rho^2]^2 H ,\nonumber
\end{eqnarray}
where we have defined the operator
\begin{equation}
\hat{T}^A := - \s^{\ga AB} \s^{\gb CD}
\rho_{BCD} {\delta \over \delta h^{\ga \gb}}
- h^{-1/2} M^{\ga \gb}  \s_{\ga}{}^{AB} \s_{\gb}{}^{CD} \rho_{BCD}.
\nonumber
\end{equation}
For $\hat{S}^A$ and $\hat{S}^A_m$, one has,
\begin{eqnarray}
\hat{S}^A \Psi_{(2)} &=& 2 \s^{\ga AB} \s^{\gb CD} \rho_{BCD}
{\delta F_1 \over \delta h^{\ga \gb } } - {2 \over 3}\beta^A
[ h^{\ga \gb} {\delta F_2 \over \delta h^{\ga \gb }} + 4 F_2 - 3 F_1 ] \} ,
\nonumber\\
\hat{S}^A \Psi_{(4)} &=&
4 [\rho^2 ] \s^{\ga AB} \s^{\gb CD} \rho_{BCD}
{\delta G_1 \over \delta h^{\ga \gb } }
+ 2 [\beta^2] \s^{\ga AB} \s^{\gb CD} \rho_{BCD}
{\delta G_2 \over \delta h^{\ga \gb } }   \nonumber \\
&+& \beta^A [\rho^2]
[ - {2\over 3} h^{\ga \gb} {\delta G_2 \over \delta h^{\ga \gb }} - {13 \over
3}
G_2 + 2 G_1 ] \} , \nonumber \\
\hat{S}^A \Psi_{(6)} &=&
4 [\rho^2 ] [\beta^2] \s^{\ga AB} \s^{\gb CD} \rho_{BCD}
{\delta H \over \delta h^{\ga \gb } } - 2 \beta^A [\rho^2]^2
[ {1\over 3} h^{\ga \gb} {\delta H \over \delta h^{\ga \gb }} + 3 H ],
\nonumber \\
\hat{S}_m^A \Psi_{(0)} &=& 3\sqrt{2}i m \beta^A E , \nonumber\\
\hat{S}_m^A \Psi_{(2)} &=& 3\sqrt{2}i m \beta^A [\rho^2] F_1 ,
 \nonumber\\
\hat{S}_m^A \Psi_{(4)} &=& 3\sqrt{2}i m \beta^A [\rho^2]^2 G_1 .
 \nonumber
\end{eqnarray}

 Our second step is to insert
these expressions in (\ref{eq:mx1}), (\ref{eq:mx2}).
Using the fact that $\beta$ and $\rho$ can be specified independently,
we arrive at the equations,
\begin{eqnarray}
h^{ab} {\delta F_1 \over \delta h^{ab}}
+ { 1 \over 6}F_2 + (-{1 \over 2} h^{-{1 \over 2}} M^{ab} h_{ab}
+ {3 \over 2} \alpha + 2 ) F_1
+ 4 \sqrt {2} im h^{-{1 \over 2}} G_2 &=& 0 ,\\
-{2 \over 3} h^{ab} {\delta G_2 \over \delta h^{ab}}
- {13 \over 3} G_2  + 2G_1
+ 3 \sqrt {2} im F_1 &=&  0 , \\
 h^{ab} {\delta G_1 \over \delta h^{ab}} + { 1 \over 6} G_2
+ (- {1 \over 2}  h^{-{1 \over 2}} M^{ab} h_{ab}
+ {3 \over 2} \alpha + 2 ) G_1
+ 4 \sqrt {2} im h^{-{1 \over 2}} H &=& 0 ,\\
-{2 \over 3} [ h^{ab} {\delta F_2 \over \delta h^{ab}}
+ 4 F_2 - 3 F_1 ] +
3 \sqrt {2} im E &=& 0 , \\
h^{ab} {\delta E \over \delta h^{ab}}
- { 1 \over 2} h^{1 \over 2} M^{ab} h_{ab} E +
 {3 \over 2} \alpha E + 4 \sqrt {2} im h^{-{1 \over 2}} F_2 &=& 0 ,
\\
-{2 \over 3} h^{ab} {\delta H \over \delta h^{ab} } -
6 H + 3 \sqrt {2} im G_1 &=& 0,
\end{eqnarray}
and
\begin{eqnarray}
\hat{T}^A f  &=&  0,
\label{eq:t1}\\
{\delta g \over \delta h^{ab} }  &\propto &  h_{ab},
\label{eq:t2}
\end{eqnarray}
with $f = E, F_1 , F_2 , G_2$, $g = F_1, G_1, G_2, H$.

For $M^{ab} \ne 0 $, using Eq. (\ref{eq:t2}) in (\ref{eq:t1}),
it follows immediately
that
\begin{equation}
 F_1 = G_2 = 0 .
\end{equation}
Using this in the remaining equations
 (21), (22), (23), (24), respectively, one finds that
\begin{equation}
F_2 = G_1 = H = E = 0.
\end{equation}

The case $M^{ab} = 0 $ (Bianchi I) must be treated separately.
However, it is easy to see that the
specialization of Eqs. (21) through (\ref{eq:t2}) gives that
$\Psi$ must vanish also for this special case.

\vskip 1em

\noindent{\bf V. Final remarks}

\vskip .5em

To conclude, we would like to compare our results with previous work.

Using the triad ADM canonical formulation, D'Eath and collaborators
have reached our same conclusion for Bianchi I and Bianchi IX
models with $\Lambda$ \cite{DEI,DEIX}. The FRW solution is
also considered in \cite{DEIX}, where a non-vanishing physical state
is found. This is a consequence of the insistence on
imposing by hand supersymmetry. This leads to throw away enough
degrees of freedom of the gravitino field that a solution survives.
In light of our result, this seems to be a pathological case. A small
amount  of anisotropy will excite the neglected gravitino
degrees of freedom, and force the wavefunction to vanish.

In the full theory with $\Lambda$,
in the connection representation, and for
a special factor ordering,  a single physical state has been proposed,
that contains an infinte number of fermionic fields \cite{FU,SANO}.
This state generalizes to supergravity the Chern-Simons type
physical state of Refs. \cite{Kodama,BGP}.
It is our hope that this paper will prompt a more systematic
investigation of such physical states.

\vskip .5em

\noindent {\bf  Acknowledgements}

\vskip .5em

R.C. thanks Jemal Guven for discussions.
O.O. was supported in part by
CONACyT grants F246-E9207 and 1683-E9209.

\bibliographystyle{plain}

\end{document}